\documentstyle[axodraw]{article}
\begin{document}
\vspace*{-1in}
\renewcommand{\thefootnote}{\fnsymbol{footnote}}
\begin{flushright}
SINP/TNP/00-21\\
\texttt{hep-ph/0007242} 
\end{flushright}
\vskip 5pt
\begin{center}
{\Large{\bf Large CP violation in radiative B decays \\
                  in supersymmetry without R-parity }}
\vskip 25pt {\sf Gautam Bhattacharyya $^{1,\!\!}$
\footnote{Electronic address: gb@tnp.saha.ernet.in}}, {\sf Darwin
Chang $^{2,\!\!}$ \footnote{Electronic address:
chang@phys.nthu.edu.tw}}, {\sf Chung-Hsien Chou $^{2,\!\!}$
\footnote{Electronic address: d833302@phys.nthu.edu.tw}}, {\sf
Wai-Yee Keung $^{3,\!\!}$ \footnote{Electronic address:
keung@uic.edu}} \vskip 10pt $^1${\it Saha Institute of Nuclear
Physics, 1/AF Bidhan Nagar, Calcutta 700064, India} \\ $^2${\it
NCTS and Physics Department, National Tsing-Hua University,
Hsinchu, 30043, Taiwan, R.O.C.} \\ $^3${\it Physics Department,
University of Illinois at Chicago, IL 60607-7059,  U.S.A.} \vskip
20pt

{\bf Abstract}
\end{center}

\begin{quotation}
{\small We demonstrate that the R-parity breaking interactions within
their current experimental upper bounds can give rise to large
mixing-induced CP asymmetry in exclusive radiative $B$ decays that may
be detectable in the upcoming $B$ factories.
\vskip 5pt \noindent
\texttt{PACS number(s)}: 11.30.Er, 13.25.Hw, 12.60.Jv, 11.30.Fs \\ 
\noindent
\texttt{Keywords:} CP violation, $B$ decays, R-parity violation}
\end{quotation}

\vskip 20pt

\setcounter{footnote}{0}
\renewcommand{\thefootnote}{\arabic{footnote}}

Supersymmetry is widely considered to be a leading candidate of
physics beyond the standard model that may be realized in Nature.  So
much so that the search for its signals constitutes one of the main
physics goals of the current and future colliders.  On the other hand,
there is an emerging consensus that new physics might show up through
detectably large CP violation in $B$ decays \cite{sm}. In this sense,
it is reasonable to ask whether it can provide an indirect but
an unmistakable signature of supersymmetry. 

\vskip 5pt

Since $B$ decay modes are often masked by theoretical as well as
experimental uncertainties, one way to establish the existence of new
physics is to identify some characteristic signatures of it which can
never be reproduced within the acceptable range of the standard model
(SM) parameters. CP asymmetry in exclusive radiative $B$ decays
offers one such sensitive probe of physics beyond the SM. More
explicitly, consider $B_q \rightarrow M^0 \gamma$, where $q$ is either
$d$ or $s$ quark, and $M = \rho^0, \omega, \phi, K^{*0}$ (decaying as
$K^{*0} \rightarrow K_s \pi^0$) is a self-conjugate meson with CP
eigenvalue $\xi = \pm 1$. The CP asymmetry would arise due to the
interference between mixing and decay. If $\phi_L$ denotes the weak
phase associated with the $b_R \rightarrow s_L \gamma_L$ (i.e. $B_q
\rightarrow M \gamma_L$) decay with amplitude $A_L$, while $\phi_R$
and $A_R$ are corresponding quantities for $b_L \rightarrow s_R
\gamma_R$ (i.e.  $B_q \rightarrow M \gamma_R$), then the
time-dependent mixing-induced CP asymmetry will be given by,
\begin{equation}
\label{acp}
a_{\rm CP} (t) \equiv \frac{\Gamma(t) - \bar{\Gamma}(t)}{\Gamma(t) +
\bar{\Gamma}(t)} = \xi A_M \sin(\phi_M - \phi_L - \phi_R) \sin(\Delta
m t),
\end{equation}
where, $A_M = 2 |A_L A_R|/(|A_L|^2 + |A_R|^2)$, and $\phi_M$ is the
phase of $B_q$-$\bar{B}_q$ mixing. In the SM, this asymmetry is small
because of the following reason. At the quark level the interference
occurs between $b_L \rightarrow q_R \gamma_R$ whose amplitude is
proportional to $m_q$ and the hermitian conjugate (through
$B_q$-$\bar{B}_q$ mixing) of $b_R \rightarrow q_L \gamma_L$ whose
amplitude is proportional to $m_b$, leading to $A_M \propto m_q/m_b$.
As a result, this asymmetry is $\sim$ 1\% for $b \rightarrow d \gamma$
and $\sim$ 10\% for $b \rightarrow s \gamma$. The reason for this
suppression is clearly the appearance of light quark mass $m_q$ as a
pre-factor with the $\bar{b}_L\sigma_{\mu \nu} q_R F^{\mu \nu}$ part
(i.e. $b_L \rightarrow q_R \gamma_R$ decay) of the effective
Hamiltonian {\em vis-a-vis} the appearance of $m_b$ with the
$\bar{b}_R\sigma_{\mu \nu} q_L F^{\mu \nu}$ part (i.e. $b_R
\rightarrow q_L \gamma_R$ decay). Is it possible to avoid this
suppression by going beyond the SM? In the left-right symmetric model
this asymmetry can go up to 50\% \cite{ags} and in supersymmetric
model to about 80\% with large sfermion mixings \cite{hou}. In this
paper, we try to answer this question in supersymmetric models {\em
without} R-parity.

\vskip 5pt

In the minimal supersymmetric standard model (MSSM), gauge invariance
ensures neither the conservation of lepton number ($L$) nor that of
baryon number ($B$). Defining $R$-parity in terms of $L$ and $B$ as $R
= (-1)^{(3B+L+2S)}$, where $S$ is the spin of the particle, one should
in a general supersymmetric model allow R-parity-violating (RPV)
couplings \cite{rpar}. $R$ is +1 for all SM particles and $-1$ for
their superpartners. Even though any concrete evidence for the
existence of RPV terms is still lacking, the recent observation of
neutrino masses and mixings in solar and atmospheric neutrino data
suggests that it would be premature to abandon the $L$-violating
interactions. However, to avoid rapid proton decay one cannot
simultaneously admit both $L$- and $B$-violating interactions and for
this reason we impose $B$ conservation by hand. The $L$-violating
superpotential can be written as (with $i, j, k$ as generation
indices)
\begin{eqnarray}
W_{\rm RPV} & \equiv &
          \frac{1}{2} \lambda_{ijk}L_i L_j E_k^c +
          \lambda_{ijk}^{\prime}L_i Q_j D_k^c
          + \mu_j L_j H_u,
\label{eq:fterm}
\end{eqnarray}
where $\lambda_{ijk} = - \lambda_{jik}$. Here $L_i$ and $Q_i$ are
lepton and quark doublet superfields, $E_i^c$ and $D_i^c$ are charged
lepton and down quark singlet superfields, and $H_u$ is the Higgs
superfield that gives mass to up-type quarks. The trilinear
$\lambda_{ijk}$-couplings and bilinear $\mu_i$ mass parameters are not
relevant for our purpose and from now on we consider only the
trilinear $\lambda'_{ijk}$-induced interactions.

\vskip 5pt

Tight constraints on the sizes of these couplings have been placed
from the consideration of neutrinoless double beta decay,
$\nu_e$-Majorana mass, charged-current universality, $e-\mu-\tau$
universality, $\nu_{\mu}$ deep-inelastic scattering, atomic parity
violation, $\tau$ decays, $D$ and $K$ decays, $Z$ decays, etc. Product
of two couplings at a time have been constrained from $\mu-e$
conversion, $\mu\rightarrow e\gamma$, $b\rightarrow s \gamma$, $B$
decays into two charged leptons, $K_L-K_S$ and $B_q-\overline{B_q}$
($q = d,s$) mass differences, etc. For a collection of all these
limits see \cite{review}.

\vskip 5pt

Let us now turn our attention to the $b \rightarrow s \gamma$
amplitude in RPV models (Needless to add that a similar analysis can
be carried out for $b \rightarrow d \gamma$) \cite{deCarlos}. The aim
is to generate unsuppressed diagrams for $b_L \rightarrow s_R
\gamma_R$. The RPV diagrams are shown in Fig.~1. The trilinear
$L$-violating couplings involved are $\lambda'_{ij2}$ and
$\lambda'_{ij3}$.

\begin{center}
\begin{picture}(150,120)(0,-20)
\ArrowLine(-4,-6)(0,0)
\Line(1,6)(7,6) \Line(4,3)(4,9)
\Line(0,0)(8,12)       \Text(2,-3)[lt]{$b_L$}
\ArrowLine(8,12)(40,60)
\ArrowLine(40,60)(72,12)
\ArrowLine(72,12)(84,-6)          \Text(80,-3)[rt]{$s_R$}
\DashArrowLine(8,12)(72,12){7}
\Photon(40,60)(40,90){4}{4} \Text(45,75)[l]{$\gamma$}
          \Text(70,30)[bl]{$d_{jL},u_{jL},\ell_{iL}$}
          \Text(40,9)[t]{$\tilde\nu_{iL},\tilde\ell_{iL},\tilde u_{jL}$}
                            \Text(40,-20)[b]{(a)}
\end{picture}
\begin{picture}(100,100)(0,-20)
\ArrowLine(-4,-6)(0,0)
\Line(1,6)(7,6) \Line(4,3)(4,9)
\Line(0,0)(8,12)       \Text(2,-3)[lt]{$b_L$}
\DashArrowLine(8,12)(40,60){7}
\DashArrowLine(40,60)(72,12){7}
\ArrowLine(72,12)(84,-6)     \Text(80,-3)[rt]{$s_R$}
\ArrowLine(8,12)(72,12)
\Photon(40,60)(40,90){4}{4} \Text(45,75)[l]{$\gamma$}
          \Text(40,9)[t]{$\nu_{iL},\ell_{iL},u_{jL}$}
          \Text(70,30)[lb]{$\tilde d_{jL},\tilde u_{jL},\tilde\ell_{iL},$}
                            \Text(40,-20)[b]{(b)}
\end{picture}
\end{center}

\begin{center}
Fig.~1: {\small \sf The diagrams for $b_L \rightarrow s_R \gamma_R$
induced by $\lambda'_{ij2}$ and $\lambda'_{ij3}$ couplings.}
\end{center}

\noindent 
(i) Figure 1a is a generic diagram in which the
photon couples to the internal fermion line. It follows from the RPV
superpotential, and as shown in the Figure 1a, that the fermion - scalar
combinations in this case are ($d_{jL}$, $\tilde{\nu}_{iL}$),
($u_{jL}$, $\tilde\ell^-_{iL}$), and ($\ell^-_{iL}$, $\tilde{u}_{jL}$).

\noindent 
(ii) Figure 1b is a similar to diagram where the photon is attached to
the internal scalar. This time the fermion - scalar combinations are
($\nu_{iL}$, $\tilde{d}_{jL}$), ($\ell^-_{iL}$, $\tilde{u}_{jL}$),
and ($u_{jL}$, $\tilde\ell^-_{iL}$).

\vskip 5pt

In Fig.~1, although chiral flips are marked only on the external $b$
quark lines thus leading to a contribution proportional to $m_b$,
similar diagrams can be drawn with chiral flips on the external $s$
quark lines as well (leading to $b_R \rightarrow s_L \gamma_L$), the
latter contribution being proportional to $m_s$. The sum of
amplitudes, taking into account all fermion-sfermion combinations in
Fig.~1, at the electroweak scale, is given by
\begin{equation}
 i \Gamma^{(1)}_\mu=-\sum_{ij}
{ie\over 32\pi^2}{\lambda'_{ij2}\lambda_{ij3}^{'*}\over m_W^2}
\bar s(m_b P_L+m_sP_R)i\sigma_{\mu\nu}q^\nu b F_R^{ij} \ ,
\end{equation}
with $F_R^{ij}=Q_d F^{ijd}+ Q_u F^{iju}+ Q_{\ell^-} F^{ij\ell^-}$, where
\begin{equation}
F^{ijd} = 
{m_W^2\over m^2_{\tilde\nu_i}}
         G_1\left({m^2_{d_j}\over m^2_{\tilde\nu_i}}\right)
+
{m_W^2\over m^2_{\tilde  d_j}}
         G_2\left({m^2_{\nu_i}\over m^2_{\tilde d_j}}\right) \ ,
        \end{equation}
\begin{equation}
F^{iju} = 
{m_W^2\over m^2_{\tilde\ell^-_i}}
         G_1\left({m^2_{u_j}\over m^2_{\tilde\ell^-_i}}\right)
+
{m_W^2\over m^2_{\tilde u_j}}
         G_2\left({m^2_{\ell^-_i}\over m^2_{\tilde u_j}}\right) \ ,
\end{equation}
\begin{equation}
F^{ij\ell^-} =
{m_W^2\over m^2_{\tilde u_j}}
         G_1\left({m^2_{\ell^-_i}\over m^2_{\tilde u_j}}\right)
+
{m_W^2\over m^2_{\tilde\ell^-_i}}
         G_2\left({m^2_{u_j}\over m^2_{\tilde\ell^-_i}}\right) \ .
\end{equation}
Here the Inami-Lim functions \cite{inami-lim} are
\begin{equation}
G_1(x)=\xi_1(x)-\xi_2(x) \ ,  G_2(x)=-\xi_0(x)+2\xi_1(x)-\xi_2(x) \ ,
\end{equation}
\begin{equation}
\xi_n(x) \equiv \int^1_0 \frac{z^{n+1}{\rm d}z}{1+(x-1)z}
=-{ {\rm ln}x+(1-x)+\cdots+{(1-x)^{n+1}\over n+1} \over (1-x)^{n+2} }
\ .
\end{equation}
The charge $Q$ factor in the expression for $F_R^{ij}$ reflects the
charge of the internal line where the photon leg is attached to.
The function $G_1$ ($G_2$) corresponds to the case that the photon
attaches to the internal fermion (sfermion).
The argument $m^2_j/m^2_{\tilde i}$ in functions $G$'s indicates
internal lines of the fermion of generation $j$ and the  sfermion of
generation $i$. Explicitly,
\begin{equation}
G_1(x) = {2+5x-x^2\over 6(1-x)^3} + \frac{x\ln x}{(1-x)^4} \ ,
\quad
G_2(x)=- {1-5x-2x^2\over 6(1-x)^3}+{x^2\ln x\over(1-x)^4} \ .
\end{equation}
Except for $j=3$ in the combination
$\lambda'_{ij2}\lambda_{ij3}^{'*}$, in which case there could be an
internal top quark line, 
we can set the argument $x\to0$ to obtain
\begin{equation}
G_1(0)={1\over3}\ , \quad G_2(0)=-{1\over6}  \ .
\end{equation}
In the limit of a common mass $\tilde m$ for all sfermions, we then have
\begin{equation}
F^{ij}_R \simeq -{m_W^2\over 9 \tilde m^2 }  \  .
\label{eq:sim}
\end{equation}
To be precise, we should treat the heavy top sector separately
according to the  detailed formulas.
However, we use Eq.~(\ref{eq:sim})  only for simplicity.

\begin{center}
\begin{picture}(150,120)(0,-20)
\ArrowLine(-4,-6)(0,0)
\Line(73,6)(79,6) \Line(76,3)(76,9)
\Line(0,0)(8,12)       \Text(2,-3)[lt]{$b_L$}
\ArrowLine(8,12)(40,60)
\ArrowLine(40,60)(72,12)
\Line(72,12)(76,6) \ArrowLine(76,6)(84,-6)          \Text(80,-3)[rt]{$s_R$}
\DashArrowLine(72,12)(8,12){7}
\Photon(40,60)(40,90){4}{4} \Text(45,75)[l]{$\gamma$}
          \Text(70,30)[bl]{$d_{jR}$}
          \Text(40,9)[t]{$\tilde\nu_{iL}$}
                            \Text(40,-20)[b]{(a)}
\end{picture}
\begin{picture}(100,100)(0,-20)
\ArrowLine(-4,-6)(0,0)
\Line(73,6)(79,6) \Line(76,3)(76,9)
\Line(0,0)(8,12)       \Text(2,-3)[lt]{$b_L$}
\DashArrowLine(8,12)(40,60){7}
\DashArrowLine(40,60)(72,12){7}
\ArrowLine(76,6)(84,-6)     \Text(80,-3)[rt]{$s_R$}
\Line(72,12)(76,6) 
\ArrowLine(72,12)(8,12)
\Photon(40,60)(40,90){4}{4} \Text(45,75)[l]{$\gamma$}
          \Text(40,9)[t]{$\nu_{iL}$}
          \Text(70,30)[lb]{$\tilde d_{jR}$}
                            \Text(40,-20)[b]{(b)}
\end{picture}
\end{center}

\begin{center}
Fig.~2: {\small \sf The diagrams for $b_L \rightarrow s_R \gamma_R$
induced by $\lambda^{\prime}_{i3j}$ and $\lambda^{\prime}_{i2j}$
couplings.}
\end{center}

In Fig.~2, we show graphs of another type, with couplings
$\lambda^{\prime}_{i3j}$ and $\lambda^{\prime}_{i2j}$. This time we do
not explicitly exhibit the diagrams where the chiral flips are on the
$b$ quark lines (corresponding to $b_R \rightarrow s_L
\gamma_L$). They give rise to the amplitude
\begin{equation}
 i \Gamma^{(2)}_\mu=-\sum_{ij}
{ie\over 32\pi^2}{\lambda'_{i3j}\lambda_{i2j}^{'*}\over m_W^2}
\bar s(m_b P_R+m_sP_L)i\sigma_{\mu\nu}q^\nu b F_L^{ij} \ ,
\end{equation}
Assuming degenerate masses for the left-handed and right-handed
squarks, we obtain $F_L^{ij}= Q_d F^{ijd}$.

\vskip 5pt

Including the two types of graphs together with the SM contribution,
we set up the effective Hamiltonian responsible to the process $b\to
s\gamma$,
\begin{equation}
{\cal H}^{\rm eff}= {1 \over 64\pi^2 m_W^2}
\sum_h \bar s\sigma_{\alpha\beta} \left(c_{7h}e
F^{\alpha\beta}+c_{8h} g_sT^aG^{a\alpha\beta} \right)
   (m_b P_{-h}+m_s P_{h}) b 
\ ,\end{equation}
with
the chirality index $h=\pm$ (or $R$, $L$) indicating
$P_\pm={1\over2}(1\pm \gamma_5)$.  The term involving $c_{8h}$
corresponds to the gluonic dipole contribution.  At the electroweak
scale $m_W$, the Wilson coefficients are given by the short-distance
amplitudes outlined above,
\begin{equation}
c_{7,8R}(m_W) = \sum_{ij}\lambda'_{ij2}\lambda'^*_{ij3} F^{ij}_{7,8R}
\ .
\end{equation}
It is straightforward to see 
$ F^{ij}_{7R}=F^{ij}_{R} $ and 
$F^{ij}_{8R}= F^{ijd}+F^{iju}$.
Also,
\begin{equation}
\label{c78mw}
c_{7,8L}(m_W) =c_{7,8L}^{\rm SM}+
\sum_{ij}\lambda'_{i3j}\lambda'^*_{i2j} F^{ij}_{7,8L}, ~~{\rm with}~~
F^{ij}_{7L}= Q_d F^{ijd},~F^{ij}_{8L}= F^{ijd}.
\end{equation}
The SM contributions are well known to be \cite{inami-lim}
(with $x_t=m_t^2/m_W^2$)
\begin{equation}
c_{7L}^{\rm SM}(m_W) = g^2 
V_{tb}V_{ts}^{*}
     x_t \left[ \frac{7-5x_t-8x_t^2}{12(1-x_t)^3}
+ \frac{2x_t-3x_t^2}{2(1-x_t)^4}\ln x_t \right] \ ,
\end{equation}
\begin{equation}
c_{8L}^{\rm SM}(m_W) = g^2 
V_{tb}V_{ts}^{*}
     x_t \left[ \frac{2+5x_t-x_t^2}{4(1-x_t)^3}
+ \frac{3x_t \ln x_t}{2(1-x_t)^4} \right]. 
\end{equation}           
For our estimation, we evaluate the Wilson coefficients at the $m_b$
scale by the simplified renormalization group evolution \cite{gsw},
\begin{equation}
c_{7h}(m_b)= \eta^{\frac{16}{23}} c_{7h}(m_W) +
\hbox{$8\over3$}
\left(\eta^{\frac{14}{23}} - \eta^{\frac{16}{23}} \right ) c_{8h}(m_W)
+N^{\rm SM}_h \ , \quad \hbox{ for }  h=L \hbox{ or } R,
\end{equation}
with  $\eta=\alpha_s(m_W)/\alpha_s(m_b)=0.56$.
The leading log QCD corrections in the SM are given by,
\begin{equation}
N_L^{\rm SM}=g^2V_{tb}V^*_{ts}
\hbox{$464\over513$} \left(\eta^{-{3\over23}}-\eta^{16\over23}\right)
\ ,\quad  N_R^{\rm SM} = 0 \ .
\end{equation}
For the purpose of {\it order-of-magnitude} estimate, we have ignored
other leading logarithmic contributions arising from additional
operators of RPV origin, which have been outlined in
Refs.~\cite{bs,cwl}.

\vskip 5pt

Numerically, for the a common mass $\tilde m$, same for squarks and
sleptons, we have
\begin{equation}
c_{7R}(m_b) =       D
        \left({m_W^2\over \tilde m^2}\right) 
        \sum_{ij} \lambda'_{ij2}\lambda_{ij3}^{'*} 
\ ,
\label{eq:D_cR}
\end{equation}
\begin{equation}
c_{7L} (m_b)=   g^2V_{tb}V_{ts}^* E  +   {1\over2}D
        \left({m_W^2\over \tilde m^2}\right) 
        \sum_{ij} \lambda'_{i3j}\lambda_{i2j}^{'*} 
\ , 
\label{eq:D_cL}
\end{equation}
for $E \simeq 0.65$ and
\begin{equation}
D=\hbox{$8\over9$} \eta^{{14\over23}}- \eta^{16\over23}
\simeq -0.044
\ .
\end{equation}

\vskip 3pt

The size of the mixing-induced CP asymmetry will be controlled by the
relative magnitude of $A_R$ and $A_L$ (for the exact calculation, one
should use Eq.~(1)). This is given by the ratio  
\begin{equation}
R  \equiv  \left|{A_R \over A_L}\right|  =
      \left|{{c_{7R} + (m_s/m_b) c_{7L}} \over 
                 c_{7L} + (m_s/m_b) c_{7R}}\right|   \ ,
\end{equation}
with the Wilson coefficients evaluated at the $m_b$ scale.

\begin{table}[ht]
\begin{center}
 $$
\begin{array}{cccc} 
\hline\hline
{ ~~\lambda'_{ij2} \lambda'_{ij3}~~} & {\rm ~~Upper~~limits~~} &
{ ~~\lambda'_{i2j} \lambda'_{i3j}~~} & {\rm ~~Upper~~limits~~} 
\\
\hline\hline
(112)(113) & 4.0\times 10^{-4} & (121)(131) & 4.0\times 10^{-4} \\
(212)(213) & 2.5\times 10^{-3} & (221)(231) & 1.1\times 10^{-2} \\
(312)(313) & 2.5\times 10^{-3} & (321)(331) & 6.1\times 10^{-2} \\
(122)(123) & 8.0\times 10^{-4} &(122)(132) & 3.3\times 10^{-3}(\dagger)\\
(222)(223) & 2.7\times 10^{-3}(*)&(222)(232) & 3.3\times 10^{-3}(\dagger)\\
(322)(323) & 2.7\times 10^{-3}(*)&(322)(332) & 3.3\times 10^{-3}(\dagger)\\
(132)(133) & 2.8\times 10^{-4} & (123)(133) & 2.8\times 10^{-5} \\
(232)(233) & 2.5\times 10^{-3}(*) &(223)(233)&3.3\times 10^{-3}(\dagger) \\
(332)(333) & 2.5\times 10^{-3}(*) &(323)(333)&3.3\times 10^{-3}(\dagger) \\
\hline\hline
\end{array}
$$
\end{center}
\caption[]{\small\sf The existing 1-$\sigma$ upper limits on the
  $\lambda'$ product
  couplings that enter into the expressions of $c_{7R}$ and
  $c_{7L}$. The meaning of $*$ and $\dagger$ symbols have been
  described in the text. A common 100 GeV mass for
  all scalars have been assumed while obtaining those limits.}
\end{table}

The existing upper limits (1-$\sigma$) on the magnitudes of
$\lambda'_{ij2} \lambda'_{ij3}$ and $\lambda'_{i2j} \lambda'_{i3j}$
combinations have been listed in Table 1 (while deriving limits the
couplings have been assumed to be real in most cases). We have assumed
a common mass of 100 GeV for whichever scalar is exchanged. In most
cases the best bounds have been obtained by multiplying the individual
upper limits on the respective couplings, listed in Ref.~\cite{wein}.
Note, $B_s$--$\bar{B}_s$ mixing (actually taking its lower limit in a
conservative sense) constrains the combinations $\lambda'_{ij2}
\lambda'_{ij3}$ via sneutrino mediated one-loop box graphs
\cite{BhRayc}. In some cases, depending on the generation indices of
the internal quarks and squarks, these bounds are stronger than those
obtained by multiplying individual bounds.  These have been marked
($*$) in Table 1. The bounds marked ($\dagger$) have been derived
\cite{cwl} from $B\rightarrow \phi K$ decays and semi-leptonic $B$
decays (Note, the authors of Ref.~\cite{cwl} have missed the color
suppression factor in the RPV amplitudes. We have included this factor
when quoting bounds in Table 1).

\vskip 5pt

As far as CP asymmetry is concerned, the size of $\sum_{ij}
\lambda^{'*}_{i2j} \lambda'_{i3j}$ is not so significant. The reason
is that this product (even set at its upper limit) multiplied by
$F_{7,8L}$ adds to a much larger contribution from $c^{\rm SM}_{7,8L}$
(see Eq.~(\ref{c78mw})).  The size of the other combination, namely
$\sum_{ij} \lambda'_{ij2} \lambda^{'*}_{ij3}$, is however crucial for
our prediction for CP asymmetry.  It follows from Table 1 that we can,
as a rough estimate, take the magnitude of this combination to be
$\sim \pm 0.01$.  This yields rather small $R \sim 3\%$. Such a tiny
effect can hardly be interpreted as a signal of new physics. On the
other hand, if one assumes that the squarks are much heavier than the
sleptons, as happens in the gauge-mediated supersymmetry breaking
models \cite{gmsb}, then in the RPV amplitudes one can consider only
the slepton-mediated graphs, as the squark-mediated ones sufficiently
decouple to leave no numerical impact. {\em For our remaining
discussions, we stick to the latter scenario}. In this case the factor
of $1\over2$ in Eq.~(\ref{eq:D_cL}) becomes 1 and the variable $D$ in
Eq.~(\ref{eq:D_cR}) is replaced by
\begin{equation}
\label{dprime}
D' = \hbox{$16\over 9$} \eta^{14\over 23} - \hbox{$3 \over 2$}
\eta^{16\over23} \simeq 0.25 \ ,
\end{equation} 
which is rather large compared to $D$. This time, with $|\sum_{ij}
\lambda'_{ij2} \lambda^{'*}_{ij3}| \sim 0.01$, we obtain $R$ as large
as 16\%.  Note that if we set this combination to be 0.04 (0.05),
which is not very unrealistic, as the limits have been set only on an
order-of-magnitude basis, $R$ may go up to $\sim 70\% ~(90\%)$, which
is a rather large effect.

\vskip 5pt 

Although in the case of CP asymmetry the impact of $\sum_{ij}
\lambda^{'*}_{i2j} \lambda'_{i3j}$ is insignificant as noted earlier,
this combination, both in magnitude and phase, may turn out to be
quite significant in the total branching ratio (BR) for $b \rightarrow
s \gamma$.  Since the BR is proportional to $(c_{7R}^2 + c_{7L}^2)$,
the RPV contribution to $c_{7L}$ in Eq.~(\ref{eq:D_cL}) interferes
with a rather large SM contribution and hence appears linearly, while
the RPV contribution to $c_{7R}$ (being the only contribution) appears
only quadratically.  Also, within $c_{7L}$, the interference between
the SM and the RPV pieces may be either constructive or destructive
depending on their relative phase. The possibility of this partial
cancellation may enable us to admit larger values of RPV couplings
required to generate a sizable CP asymmetry, while stay perfectly
consistent with the BR constraints. In Fig.~3, as an illustrative
example. we exhibit our prediction for the BR as a function of
$|\lambda^2|$, where $\lambda^2\equiv \sum_{ij} \lambda'_{ij2}
\lambda^{'*}_{ij3} =\sum_{ij} \lambda'_{i3j} \lambda^{'*}_{i2j}$. The
different angles refer to the relative phase between the two pieces in
Eq.~(\ref{eq:D_cL}). In Fig.~4, we demonstrate how large CP asymmetry
(only its absolute value) can be generated as a function of
$|\lambda^2|$ without violating the BR constraint. For the latter, we
used Eq.~(1) with the sine of the phase combination set to unity. Here
we wish to make the following remark. While drawing the new physics
curves in Fig.~3, we took a conservative approach that the SM
reference line is at its present mean value. One might as well place
the SM reference line close to its lower limit so that larger values
of $|\lambda^2|$ could be allowed leading to a possibility of larger
CP asymmetry.

\vskip 5pt

Before closing, we highlight the following salient features: 
\begin{enumerate}
\item The sign of the RPV induced CP asymmetry is arbitrary, since the
signs of the $\lambda'$ couplings are {\em a priori} unknown. It is
therefore possible to have a CP asymmetry not only large compared to
the SM prediction but also with an opposite sign. The occurance of a
sign flip, in particular, may constitute an unmistakable signature of
new physics \cite{ali}. 

\item In principle, the $\lambda'$ couplings need not be complex in
order to generate CP asymmetry. More precisely, in Eq.~(1) even with
$\phi_R = 0$ (i.e. with real $\lambda'$ couplings), one may obtain a
non-zero sine function from the $B$-$\bar B$ mixing phase $\phi_M$ or
the SM decay phase $\phi_L$. Nevertheless, a non-zero $\phi_R$ may
trigger a much larger CP violating effect.

\item Here we present a simple way of understanding how large CP
asymmetry can one predict without violating the branching ratio
constraint. The present experimental BR on $B \rightarrow X_s \gamma$
is $(3.15\pm 0.54) \times 10^{-4}$ \cite{cleo}, while the present SM
next-to-leading order estimate for BR ($b \rightarrow s \gamma$) is
$(3.28 \pm 0.33)\times 10^{-4}$ \cite{cm}. To check whether our
prediction of a large CP asymmetry at all contradicts the constraints
on the BR, let us consider the following illustrative example. Assume
that new physics contributes only to $c_{7R}$. In our case, this
amounts to saying that $\sum_{ij} \lambda'_{ij2} \lambda^{'*}_{ij3}$
is non-zero, while $\sum_{ij} \lambda'_{i3j} \lambda^{'*}_{i2j}$ is
vanishingly small. Recall, to a good approximation, $c_{7R}/c_{7L}
\equiv \alpha$ is a quantity that controls the size of the CP
asymmetry (for an exact expression, see Eq.~(1)). Then the BR, being
proportional to $(c_{7R}^2 + c_{7L}^2)$, is modified by an overall
factor $(1 + \alpha^2)$. Now to demonstrate how large $\alpha$ one can
tolerate, let us imagine a situation when the SM prediction of the BR
is at its present 90\% CL lower limit, i.e. $2.74 \times 10^{-4}$, and
the experimental BR settles at its 90\% CL upper limit, i.e. $4.04
\times 10^{-4}$. This means that within the current experimental and
theoretical uncertainties a 50\% $(= \alpha^2)$ new physics induced
enhancement of BR is perfectly tolerable, and hence a CP asymmetry as
large as $2\alpha/(1+\alpha^2) \sim$ 100\% (putting $\alpha = 0.7$ in
Eq.~(1)) is very much consistent with the present constraints on the
BR at 90\% CL.

\item The RPV models are better placed than the left-right symmetric
model in generating large CP asymmetry. The reason is that the
R-parity conserving MSSM contribution to $c_{7L}$ may interfere
destructively with the SM amplitude in some region of the parameter
space \cite{masiero}. Just like the RPV contribution to $c_{7L}$, this
also helps to keep the BR under control while admitting large RPV
couplings responsible for large CP asymmetry. More specifically, the
charged Higgs (chargino) loops interfere constructively
(destructively) with the SM $W$ loops, and we know that in the exact
supersymmetric limit the net amplitude vanishes \cite{bg}.

\item The combinations $\lambda'_{i12}\lambda^{'*}_{i13}$, within
their current experimental bounds, not only induce large CP asymmetry
in $b \rightarrow s \gamma$ decay, but also enhance the direct CP
asymmetry in $B^\pm \rightarrow \pi^\pm K$ channels \cite{bd}. Hence
if upcoming $B$ factories indicate large CP asymmetry in both these
(uncorrelated) channels, and perhaps with signs opposite to their
respective SM predictions, it may constitute an unambiguous signal of
the RPV scenario. We must admit though that the extraction of $\gamma$
from $B \rightarrow \pi K$ would be even more difficult in the
presence of new physics contamination.
\end{enumerate}

\vskip 10pt 
\noindent 
GB thanks the following centers for hospitality where portions of the
work were done: NCTS, Hsinchu, Taiwan; CERN Theory Division, Geneva,
Switzerland; and MPI Kernphysik, Heidelberg, Germany.  DC wishes to
thank SLAC Theory Group and U.S. DOE for hospitality while revising the
final manuscript.

\newpage \epsffile[70 95 519 583]{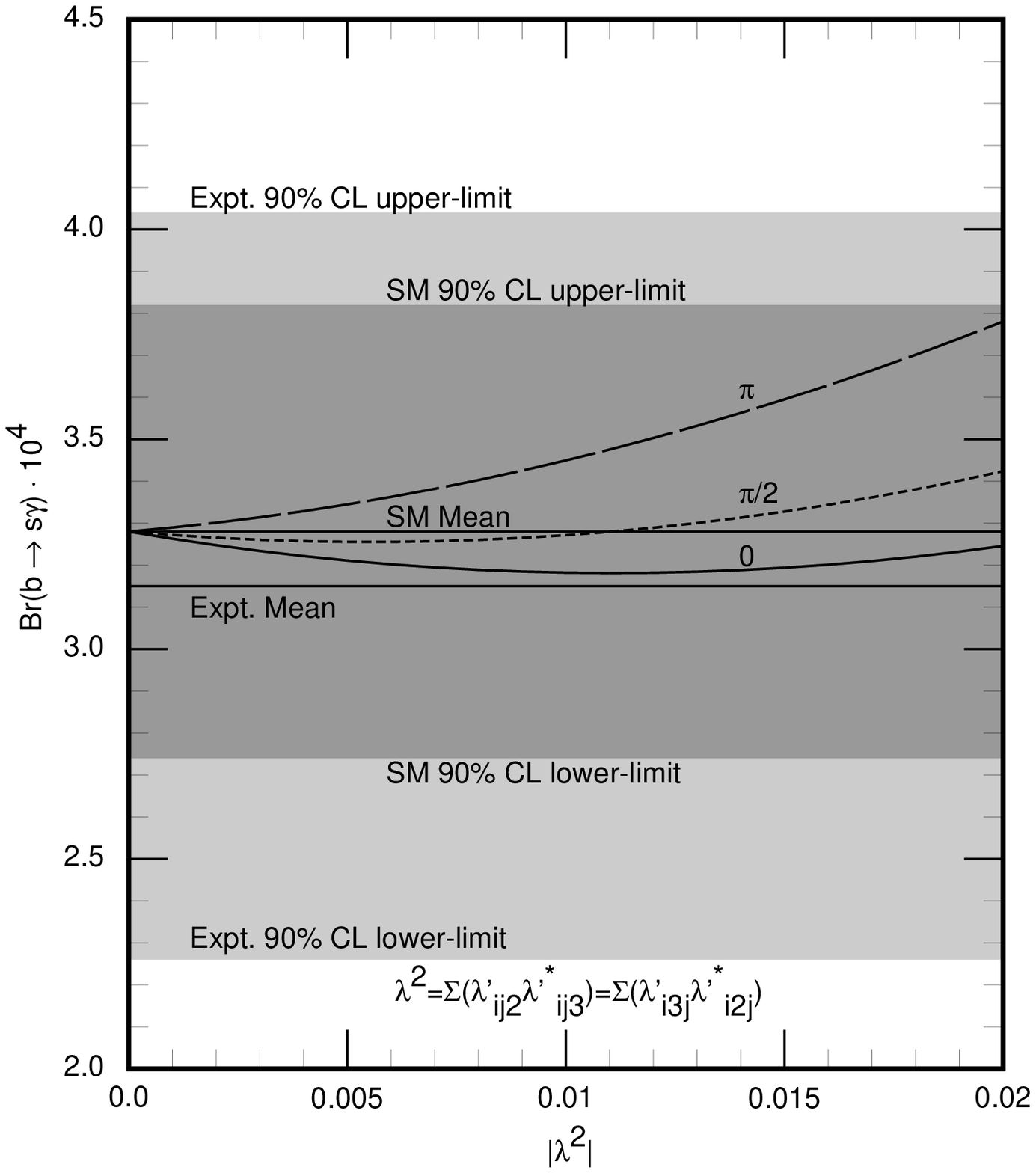}

\noindent
{\small Fig.~3: \sf Br($b\rightarrow s\gamma$) versus $|\lambda^2|$
for various choices of the relative phase between $\sum_{ij}
\lambda'_{i3j} \lambda^{'*}_{i2j}$ and $V_{tb}V_{ts}^*$. The SM
prediction and the experimentally allowed region are shaded.}

\newpage \epsffile[70 95 519 583]{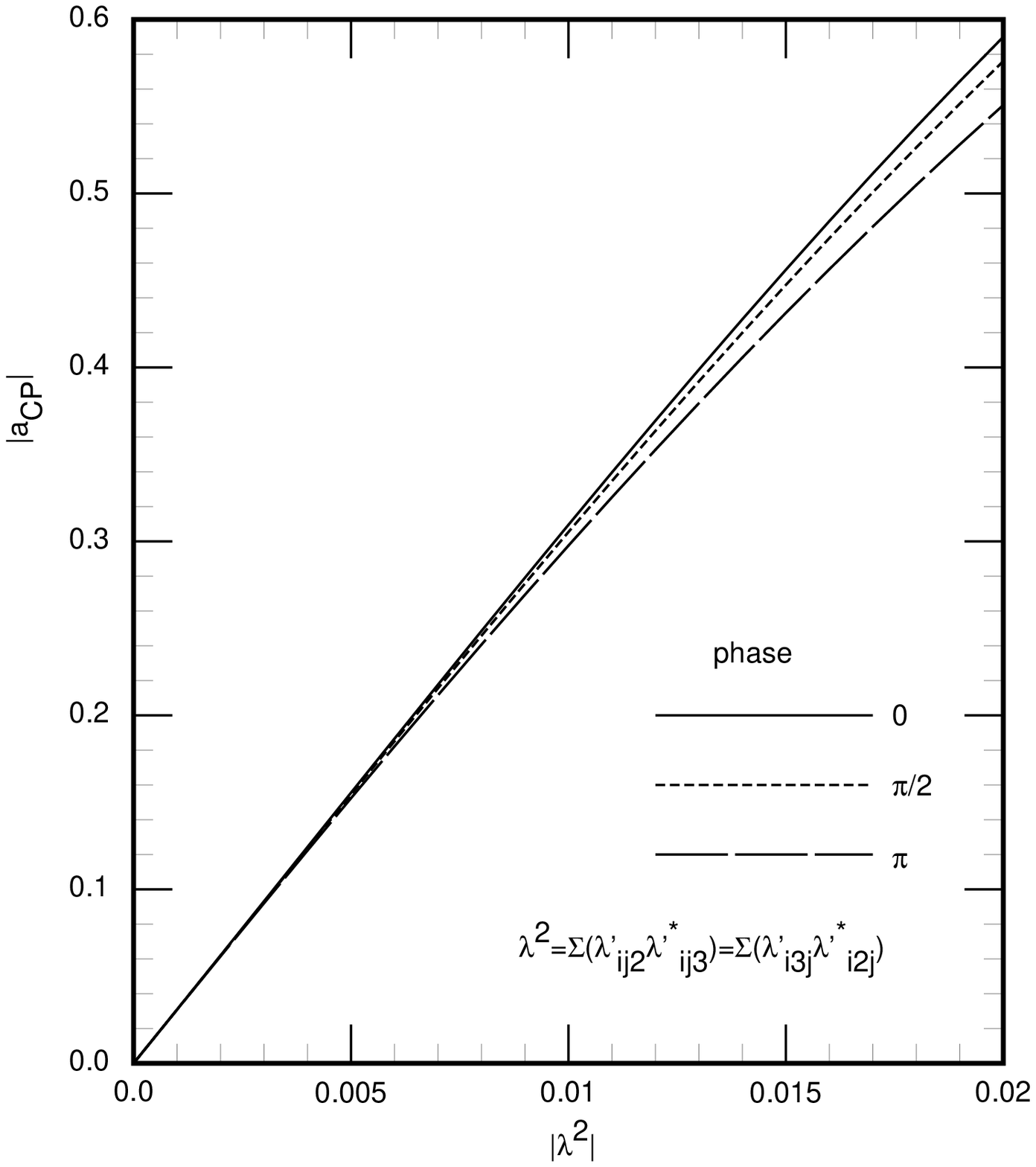}

\noindent
{\small Fig.~4: \sf The magnitude of the CP asymmetry for various
choices of the relative phase between $\sum_{ij} \lambda'_{i3j}
\lambda^{'*}_{i2j}$ and $V_{tb}V_{ts}^*$. Here we assume that both the
sine factors are unity in Eq.~(1).}


\newpage

\begin{thebibliography}{99}

\bibitem{sm} See, for example, A. Buras and R. Fleischer, in `Heavy
Flavours II', {\em ed}: A. Buras and M. Linder (World Scientific,
1997), pp.~65-238.

\bibitem{ags} D. Atwood, M. Gronau and A. Soni, Phys. Rev. Lett. 79,
185 (1997) [hep-ph/9704272]. See also, Y.-L. Wu, Chin. Phys. Lett. 16,
339 (1999) [hep-ph/9805439]. 

\bibitem{hou} C.K. Chua, X.G. He and W.S. Hou, Phys. Rev. D60, 014003
(1999) [hep-ph/9808431].

\bibitem{rpar} G. Farrar and P. Fayet, Phys. Lett. B 76, 575 (1978);
S. Weinberg, Phys. Rev. D 26, 287 (1982);
N. Sakai and T. Yanagida, Nucl. Phys. B 197, 533 (1982);
C. Aulakh and R. Mohapatra, Phys. Lett. B 119, 136 (1982).


\bibitem{review} For recent reviews, see G. Bhattacharyya, Nucl.
  Phys. B (Proc. Suppl.) 52A, 83 (1997) [hep-ph/9608415];
  hep-ph/9709395, Invited talk given at Workshop on Physics Beyond the
  Standard Model: Beyond the Desert: Accelerator and Nonaccelerator
  Approaches, Tegernsee, Germany, 8-14 Jun 1997 (and its update in the
  Proc. of the WEIN98 Symposium, Santa Fe, USA, June 1998, p. 60-69;
  eds: C. Hoffman, P.  Herczeg, H.V. Klapdor-Kleingrothaus; World
  Scientific, Singapore, 1999); H. Dreiner, hep-ph/9707435, published
  in `Perspectives on Supersymmetry', Ed. by G.L. Kane, World
  Scientific, Singapore; R.  Barbier {\it et al.}, Report of the Group
  on $R$-parity violation, hep-ph/9810232.

\bibitem{deCarlos} B. de Carlos and P.L. White, Phys. Rev. D 55, 4222
(1997) [hep-ph/9609443]. 

\bibitem{inami-lim} T. Enami and C.S. Lim, Prog. Theor. Phys. 65, 297
(1981). 

\bibitem{gsw} B. Grinstein, R. Springer and M. Wise, Nucl. Phys. B 339,
269 (1990), see their Eqs.~(3.6) and (3.86). See also, M. Ciuchini
{\em et al.}, Phys. Lett. B 316, 127 (1993); M. Misiak, Phys. Lett. B
321, 113 (1994); G. Cella {\em et al.}, Phys. Lett. B 325, 227 (1994).

\bibitem{bs} Th. Besmer and A. Steffen, hep-ph/0004067.
  
\bibitem{cwl} E.J. Chun, K. Hwang and J.S. Lee, hep-ph/0005013. This
paper contains a discussion on direct and mixing-induced CP asymmetry.

\bibitem{wein} We have used the bounds listed in the talk by
  Bhattacharyya in the Proc. of International WEIN98 Conference
  \cite{review}. These bounds are in close agreement to those in
  B.C. Allanch, A. Dedes and H. Dreiner, Phys. Rev. D 60, 075014
  (1999). Also see, O. Lebedev, W. Loinaz and T. Takeuchi,
  Phys. Rev. D 62, 015003 (2000), for updated bounds fom LEP
  observables.

\bibitem{BhRayc} G. Bhattacharyya and A. Raychaudhuri, Phys. Rev. D
  57, 3837 (1998) [hep-ph/9712245].  

\bibitem{gmsb} G.F. Giudice and R. Rattazzi, Phys. Rept. 322, 419
(1999); {\em ibid.} 322, 501 (1999) [hep-ph/9801271].

\bibitem{ali} The possibility of a sign flip in the CP asymmetry in
the $B \rightarrow \rho \gamma$ channel in the R-parity conserving
MSSM framework has recently been discussed in: A. Ali, L.T. Handoko
and D. London, hep-ph/0006175. For a discussion on CP asymmetry in
inclusive $b \rightarrow s \gamma$ as a probe of new physics, see:
A.L. Kagan and M. Neubert, Phys. Rev. D 58, 094012 (1998)
[hep-ph/9803368]; K. Kiers, A. Soni and G.-H. Wu, hep-ph/0006280.

\bibitem{cleo} S. Ahmet {\em et al.}, CLEO-CONF 99-10 [hep-ex/9908022]. 

\bibitem {cm} A. Czarnecki and W.J. Marciano, Phys. Rev. Lett. 81, 277
(1998) [hep-ph/9804252]; K. Chetyrkin, M. Misiak and M. M\"{u}nz,
Phys. Lett. B 400, 206 (1997); Erratum {\em ibid.} B 425, 414 (1998)
[hep-ph/9612313]. 

\bibitem{masiero} F. Gabbiani, E. Gabrielli, A. Masiero and
L. Silvestrini, Nucl. Phys. B 477, 321 (1996) [hep-ph/9604387]. 

\bibitem{bg} R. Barbieri and G. Giudice, Phys. Lett. B 309, 86 (1993)
[hep-ph/9303270].

\bibitem{bd} G. Bhattacharyya and A. Datta, Phys. Rev. Lett. 83, 2300
(1999) [hep-ph/9903490].

\end{thebibliography}
\end{document}